\documentclass[a4paper]{article}

\usepackage{INTERSPEECH2020}
\usepackage{multirow}
\usepackage{graphicx}
\usepackage{amsmath,amssymb}
\usepackage{microtype}
\usepackage{subfigure}
\usepackage{cite}
\usepackage{url}
\usepackage{xcolor}
\usepackage[capitalize,nameinlink]{cleveref}

\newcommand{\bw}{\mathbf{w}}

\newcommand{\bee}{\begin{eqnarray}}
\newcommand{\eee}{\end{eqnarray}}

\newlength{\widebarargwidth}
\newlength{\widebarargheight}
\newlength{\widebarargdepth}

\newcommand{\eat}[1]{}

\newcommand{\btx}{\tilde{\mathbf{x}}}
\newcommand{\bty}{\tilde{y}}

\newcommand{\bx}{\mathbf{x}}

\newcommand{\by}{\mathbf{y}}

\providecommand*\email[1]{\href{mailto:#1}{#1}}

\usepackage{pifont}
\newcommand{\cmark}{\text{\ding{51}}}
\newcommand{\xmark}{\text{\ding{55}}}

\title{Meta-Learning for Short Utterance Speaker Recognition with\\ Imbalance Length Pairs}
\name{Seong Min Kye$^1$, Youngmoon Jung$^1$, Hae Beom Lee$^1$, Sung Ju Hwang$^{1,2}$, Hoirin Kim$^1$}

\address{
  $^1$KAIST, South Korea, $^2$AITRICS, South Korea}
\email{\{kye9165,dudans,haebeom.lee,sjhwang82,hoirkim\}@kaist.ac.kr}

\begin{document}

\maketitle

\begin{abstract}
In practical settings, a speaker recognition system needs to identify a speaker given a short utterance, while the enrollment utterance may be relatively long. However, existing speaker recognition models perform poorly with such short utterances. To solve this problem, we introduce a meta-learning framework for imbalance length pairs. Specifically, we use a Prototypical Networks and train it with a support set of long utterances and a query set of short utterances of varying lengths. Further, since optimizing only for the classes in the given episode may be insufficient for learning discriminative embeddings for unseen classes, we additionally enforce the model to classify both the support and the query set against the entire set of classes in the training set. By combining these two learning schemes, our model outperforms existing state-of-the-art speaker verification models learned with a standard supervised learning framework on short utterance (1-2 seconds) on the VoxCeleb datasets. We also validate our proposed model for unseen speaker identification, on which it also achieves significant performance gains over the existing approaches. The codes are available at \textcolor{magenta}{\url{https://github.com/seongmin-kye/meta-SR}}.
\end{abstract}

\section{Introduction}
Speaker recognition (SR) with short utterance is an important challenge since in practical settings, where test utterance could be as short as $1$ to $5$ seconds. In the past decades, i-vector \cite{dehak2010front} combined with probabilistic linear discriminant analysis (PLDA) \cite{garcia2011analysis, prince2007probabilistic} has been a dominant approach for SR.  However, recently, deep neural network (DNN) based methods have shown to outperform i-vector based systems, achieving better performance on short utterances \cite{bhattacharya2017deep} over non-deep learning counterparts. Although the recent advances in deep learning make it possible to obtain impressive performance on SR such as speaker verification (SV) and identification (SI), obtaining sufficiently high performances under practical settings (e.g. short duration SR, unseen speaker SI) remains as a challenging problem. Some recent works propose to tackle these issues. Regarding SR with short utterances, \cite{gao2018improved, hajavi2019deep, MSA} introduce task-specific feature extractor which allows to extract more information from short utterances. \cite{xie2019utterance, hajavi2019deep, MSA} introduce aggregation methods to attend to more informative frames in frame-level features. In addition to these approaches, various attempts have been made to deal with short utterances. However, they do not provide a substantial performance improvements under realistic scenarios.

In this work, we aim to tackle this problem by meta-learning with imbalance length pairs. Specifically, we organize each episode such that it contains a support set of long utterances and a query set of variable short utterances. By optimizing sequence of episodes, we can train our network to match long-short utterance pairs well over conventional (referred to as ‘vanilla’) training, which optimizes for the same-length utterances. Also, variable length of query utterances allows the model to consider various practical situations during meta-training and thus allows us to obtain a more length-robust model for SR.

\begin{figure}[t!]
	\vspace{-0.45in}
	\centering
	\hfill
	\subfigure[Vanilla training]{\includegraphics[clip,
	trim=0.5cm 0.9cm 0.5cm 0.3cm,
	width=3.9cm]{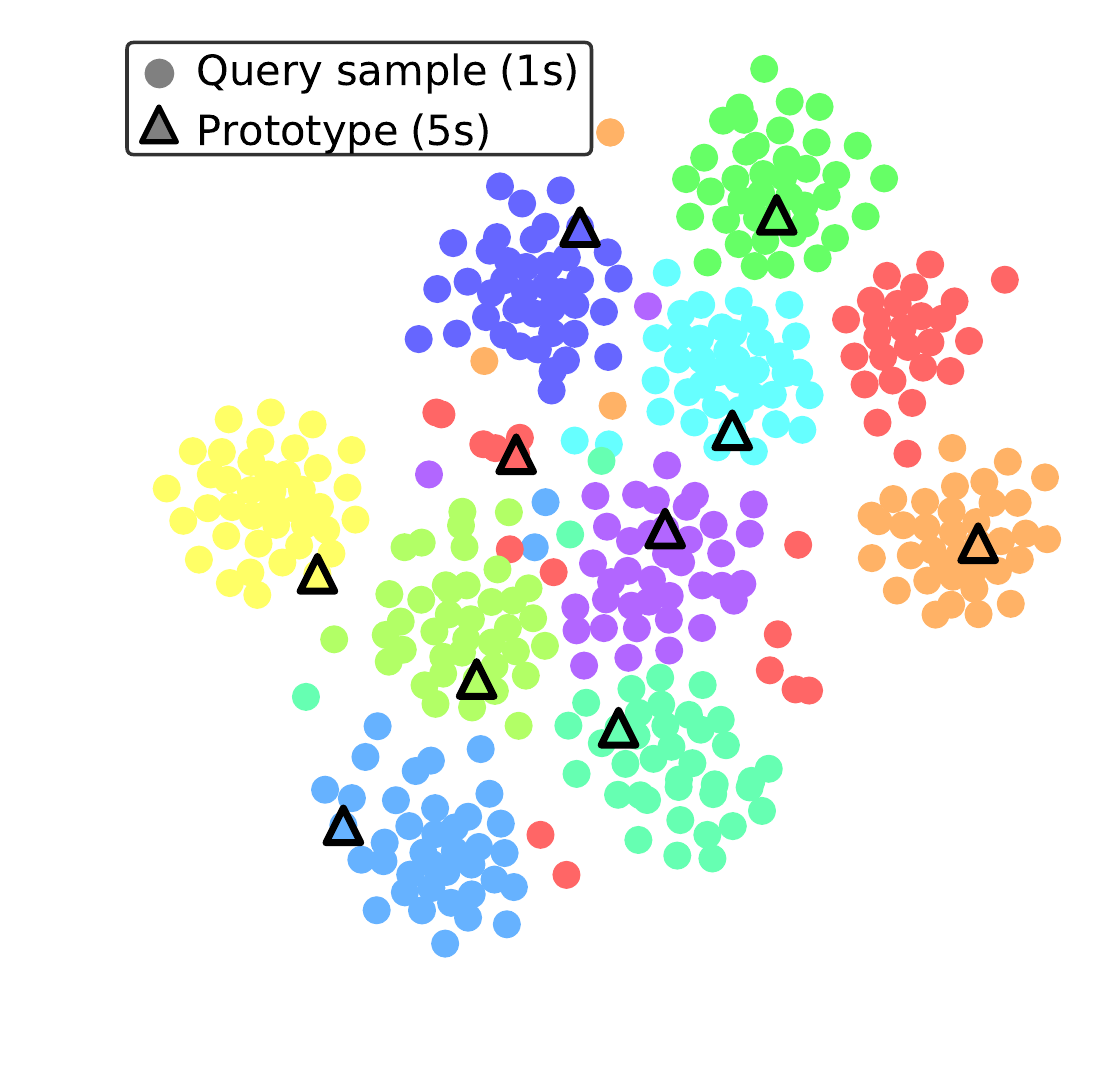}}
    \hfill
	\subfigure[Ours]{\includegraphics[clip,
	trim=0.5cm 0.9cm 0.5cm 0.3cm, width=3.9cm]{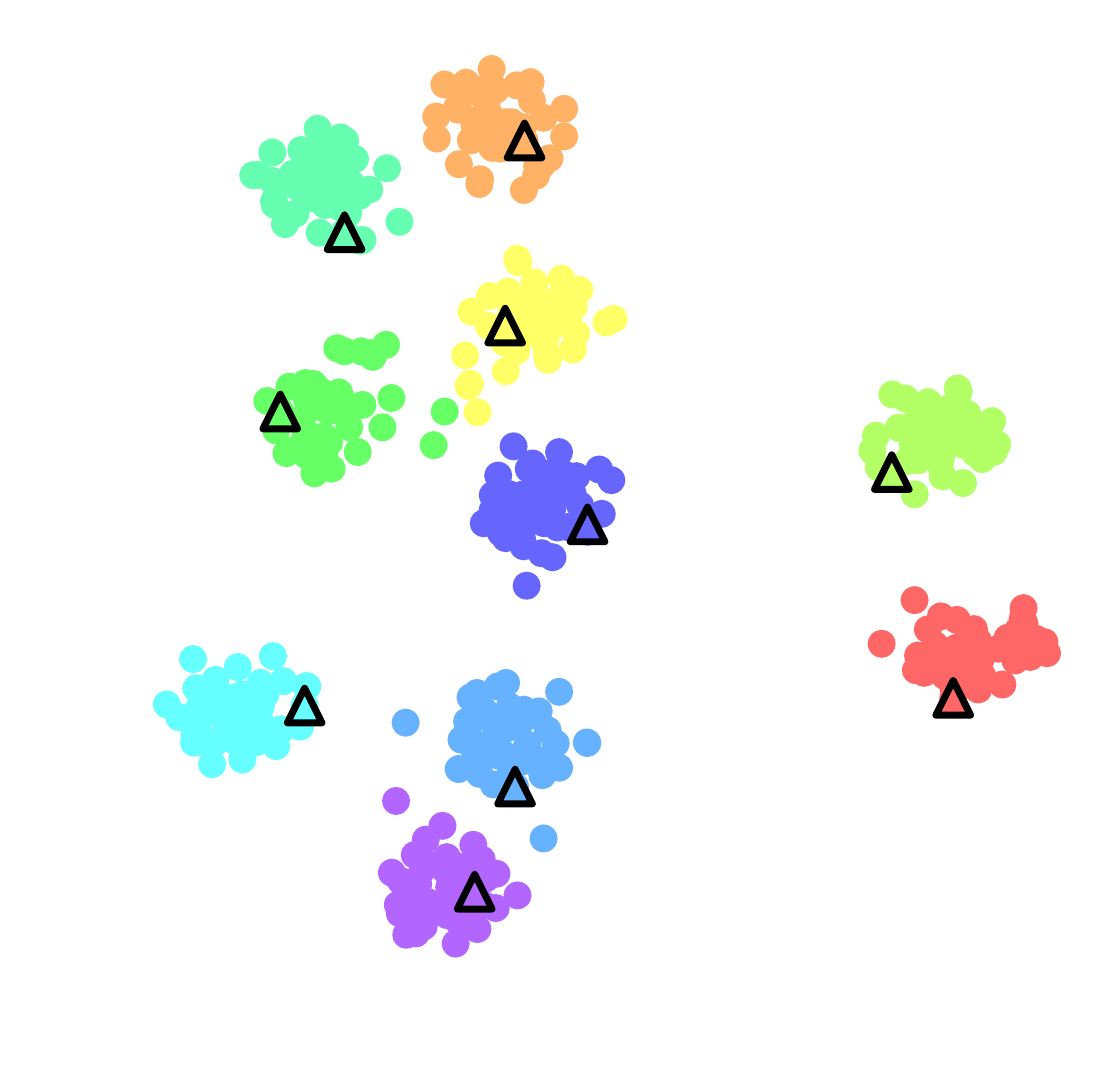}}
 	\vspace{-0.15in}
	\hfill
    \caption{
    Comparison between (a) vanilla training and (b) meta-learning with global classification. We visualize t-SNE embeddings on 10-way 5-shot task. Prototype is average of support set which consists of 5 seconds utterances. Best viewed in color.}
	\vspace{-0.25in}
    \label{fig:compare_training}
\end{figure}

\begin{figure}[t!]
\centering
\vspace{-0.45in}
\includegraphics[width=0.90\linewidth]{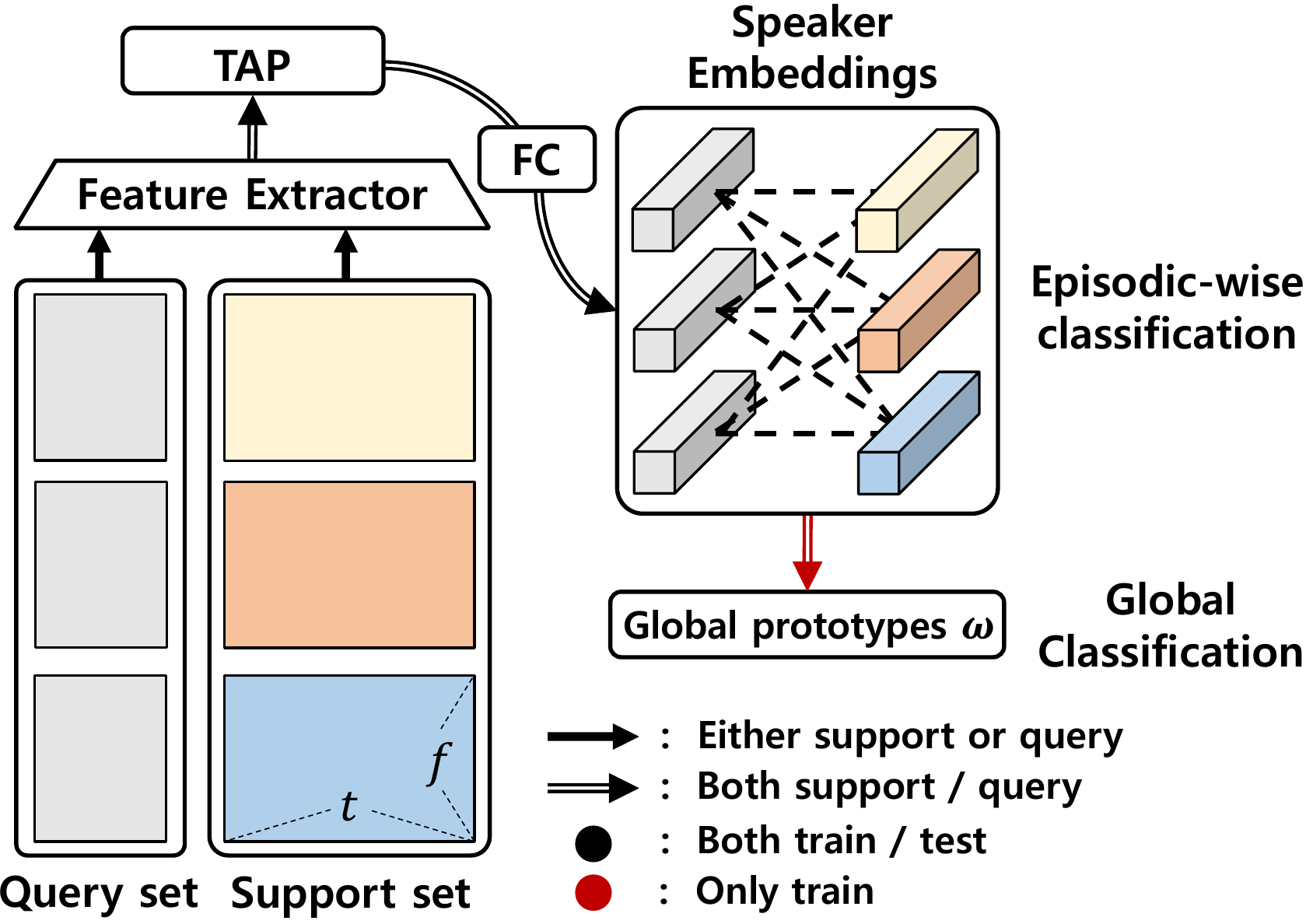}
\vspace{-0.05in}
\caption{Overview of proposed meta-learning scheme on 3-way 1-shot task. We denote the features for each speaker using different colors.}
\vspace{-0.24in}
\label{fig:network}
\end{figure}

Yet, a crucial problem here is that the query samples may not be discriminative against the entire set of classes (speakers) in the training set. Thus, we further classify every sample in each episode against the whole training classes (referred to as ‘global classification’). In doing so, an embedding of a short utterance becomes discriminative against other classes and is matched to its own long utterance at the same time (See Figure \ref{fig:compare_training}). Also, having a consistent framework across training and test phases, by targeting unseen speakers during training, allows the model to verify and identify unseen speakers well (See Figure \ref{fig:network}).


Our proposed learning scheme uses a ResNet34 \cite{he2016deep} as the base network architecture, which is widely used for SR. To verify the efficacy of our proposed model, we use a simplistic framework for implementation of our model, using naive pooling methods such as Temporal Average Pooling (TAP) for aggregation and non-margin metric loss. Also, we use 40-dimensional log mel-filterbank features as inputs to reduce time complexity, since the execution time should be short in realistic settings. We experiment on various settings, such as short utterance SV and unseen SI including conventional experimental settings (full utterance SV). We use VoxCeleb datasets \cite{nagrani2017voxceleb, chung2018voxceleb2} to directly compare with other models. Our model obtains the state-of-the-art results on short utterance SV and unseen SI on various datasets, although we use simpler implementations with smaller-dimensional features.  

Our main contributions are as follows:
\vspace{-0.03in}
\begin{itemize}

\item We propose a meta-learning framework for short utterance speaker recognition, in which each episode is composed of support and query pairs with imbalance length utterances.


\item We further propose a training objective that combines the episodic classification loss with the global classification loss, which allows to obtain well-matched and discriminative embeddings.

\item We validate our model on VoxCeleb datasets under various realistic scenarios, including speaker verification with short duration and unseen speaker identification, and achieve the state-of-the-art results.
\end{itemize}

\section{Related Work}

\textbf{DNN based speaker embedding:} Recently, DNN based methods \cite{variani2014deep, li2017deep, snyder2018x, zhang2018text, disentangle, MIRNet}, have achieved impressive performance on speaker recognition (SR), outperforming traditional i-vector systems. The key components of DNN based systems are feature extractors, aggregation of temporal features and optimization. First, many SR systems use 1D or 2D convolutional neural networks and recurrent neural networks as feature extractors, which make it possible to extract the time and frequency properties of the speaker features (MFCC, mel-filterbank). Then these extracted frame-level features are summarized into fixed-length vectors using aggregation methods which aims to capture intrinsic speaker information, such as attentive statistic pooling (ASP) \cite{okabe2018attentive}, self attentive pooling (SAP) \cite{cai2018exploring}, learnable dictionary encoding (LDE) \cite{cai2018exploring} and spatial pyramid encoding (SPE) \cite{jung2019spatial}. Then, these utterance-level features are used as inputs for with softmax classifiers with fully-connected layers. However, since softmax classifier may not obtain sufficiently discriminative embedding spaces, recent methods such as A-softmax \cite{liu2017sphereface}, AM-softmax \cite{wang2018additive} and AMM-softmax \cite{deng2019arcface} propose angular margin-based metric to reduce per-class variance.

\textbf{Metric-based meta-learning for few-shot classification:} Speaker verification and speaker identification for unseen speaker are essentially few-shot learning tasks. The goal of few-shot classification is to correctly classify unlabeled query (test) examples with only a few labeled support (enrollment) set per class. Since labeled data is scarce, conventional supervised learning in this case is prone to overfitting, and thus recent works resort to \emph{meta-learning}, which learns a model that generalize over diverse tasks to consider larger number of problems for training. One of the most popular meta-learning approach is metric-based meta-learning \cite{kye2020transductive, vinyals2016matching, snell2017prototypical, liu2019fewTPN}, which learns an embedding space that is learned to be discriminative for any given tasks. Several recent works~\cite{wang2019centroid, anand2019few} use Prototypical Networks~\cite{snell2017prototypical}, which is such a metric-based meta-learning model, for speaker recognition. Compared with above methods, our algorithm uses not only Prototypical Networks with imbalance length pairs but also global classification over the whole samples in episode.

\textbf{Speaker recognition for short utterance:} Speaker recognition for short utterance is especially challenging since the input data contains very little information about the speaker. To tackle this problem, \cite{gao2018improved, hajavi2019deep, xie2019utterance, MSA} propose aggregation techniques to extract as much information as possible from short speech. NetVLAD / GhostVLAD \cite{xie2019utterance} use attention-based pooling with learnable dictionary encoding, and time-distributed voting (TDV) \cite{hajavi2019deep} utilizes short-cut connection information with weighted sum of them, which yields impressive performance gains over GhostVLAD for short utterance SR. However, TDV obtains good performance only on short utterances and relatively low performance on long utterances. Other approaches to tackle the short utterance SR use knowledge distillation \cite{jung2019short}, generative adversarial networks \cite{zhang2018vector} and angular margin-based method \cite{huang2018angular, gusev2020deep}.
\section{Method}
In this work, we consider a practical setting for unseen speaker recognition, where the length of test utterance is shorter than the enrollment utterance. To solve this problem, the model should not only match a pair of utterances with different lengths from the same speaker, but also be able to recognize unseen speakers not included in training set well. To this end, we introduce a metric-based meta-learning scheme, in which support and query sets consist of long and short utterances, respectively. We also classify both support and query sets over the entire set of training classes rather than training them only for the given set of classes to obtain even more discriminative embeddings. 

\subsection{Problem Definition}
The goal of few-shot unseen short utterance speaker recognition problem is to recognize test utterance $\btx_i$ which is as short as 5 seconds as $i$-th speaker, given only few utterances $x_i$ from each speaker for enrollment, whose lengths could be longer than 5 seconds. Since the number of enrollment examples per class is too small, conventional supervised learning may obtain suboptimal performance due to overfitting.  

Thus, we tackle the problem with episodic meta-learning where we learn a model over diverse tasks, such that the model learns to recognize the speaker for \emph{any} utterances, while considering a different classification problem at each time. To this end, we first compose task episodes with support set and query set. We first randomly sample $N$ classes from the given dataset, and then sample $K$ and $M$ examples from each class as the support set and query set, respectively. We define the task sampling distribution as $p(\tau)$. As a result, we have a support set $\mathcal{S} = \{(\bx_i,y_i)\}_{i=1}^{N \times K}$ and a query set $\mathcal{Q} = \{(\btx_i,\bty_i)\}_{i=1}^{N \times M}$, where $y,\bty \in \{1,\dots,N\}$ are the class labels.


\subsection{Meta-learning with imbalance length pairs}
Despite large progress on speaker recognition, speaker recognition with short utterances remains to be very challenging in realistic settings due to length mismatch between training and test utterances. As shown in \cite{kanagasundaram2019study}, in conventional training setting, this can result in performance degeneration for short utterances. A prior work~\cite{gusev2020deep} tackles this problem by training the model with short segments only, which outperforms models with long utterances, but the model obtains relatively poor performances on long speech. This trade-off is problematic in realistic settings where the test utterances could be given in any lengths.

How can we then train a length-robust model for speaker recognition? To tackle this problem, we construct the training episodes for our meta-learning framework, to contain imbalanced length pairs. In practical settings, while we can enroll long utterances to the system, the test utterances could come in any variable lengths. To simulate this situation in training phase, we set the length of the utterance for the support set to be longer than the query utterances. On the other hand, we construct the query sets such that they have variable lengths (1 to 2 seconds), that are smaller than the lengths of the support utterances. 

As with \cite{snell2017prototypical}, we compute the class prototypes by averaging over the support set and enforce the query examples to become closer to their own prototypes. First, we define $\mathcal{S}_c$ as the set of support examples in class $c$ and then compute the prototype of each class $c = 1,\dots,N$ in episode:
\begin{align}
P_c = \frac{1}{|\mathcal{S}_c|} \sum_{x \in \mathcal{S}_c} f_\theta(\bx)
\end{align}
\noindent Then, we compute the distance between each query and its corresponding prototype. In this work, we use cosine similarity as the distance metric:
\begin{align}
    &d(f_\theta(\btx_i), P_c) = \frac{f_\theta(\btx) \cdot P_c}{\|P_c\|_2} =\|f_\theta(\btx)\|_2 \cdot cos(\theta_{i,c})
    \label{eq:dist_metric}
\end{align}
\noindent where it can be seen as the cosine similarity with an input-wise length scale. We can then obtain the probability of the sample belonging to each class $c$ as follows:
\begin{align}
p(\bty=c|\btx,\mathcal{S};\theta) = 
\frac{\exp(d(f_\theta(\btx), P_c))}
{\sum_{c'=1}^C \exp(d(f_\theta(\btx), P_{c'}))}
\label{eq:conf}
\end{align}
\noindent Then, we compute the loss for each episode:
\begin{equation}
L_e^\tau(\theta) = \frac{1}{|Q|}\sum_{(\btx,\bty) \in Q} -\log p(\bty|\btx,S;\theta)
\label{eq:episode_loss}
\end{equation}
\subsection{Global classification}
With the proposed meta-learning scheme, we can make variable short utterance to be close to relatively long utterance. However, optimization only within each episode may fail to learn a discriminative embedding space. Inspired by \cite{kye2020transductive}, we additionally classify support and query samples against whole training classes. By optimizing support and query samples of different length at once, we can reduce per class variance according to utterance duration. At the same time, we can make the discriminative embeddings over all other classes in training set. Following ~\cite{kye2020transductive}, we assume a set of global prototypes for each class:
\begin{align}
\omega = \{\bw_c \in \mathbb{R}^l|c=1,\dots,C'\}
\label{eq:global_prototype}
\end{align}
\noindent where $C'$ is the number of classes in entire training set and $l$ is dimension of embedding. For $(\bx, \by) \in S \cup Q$, we predict the probability of the sample $x$ being an instance of class $y$ as follows:
\begin{equation}
p(\by|\bx;\theta,\omega) =
\frac{\exp(d(f_\theta(\bx), {\bw_{\by}}))}
{\sum_{c=1}^{C'} \exp(d(f_\theta(\bx), {\bw_{c}}))}
\label{eq:pixel_pred}
\end{equation}
and compute the global loss:
\begin{equation}
L_g^\tau(\theta,\omega) = \frac{1}{|S|+|Q|}\sum_{(\bx,\by) \in S \cup Q} -\log p(\by|\bx;\theta,\omega)
\label{eq:global_loss}
\end{equation}
\noindent where $d$ is the distance metric described in Eq. 2. Note that the global classification is conducted on both the support and query samples. Finally, our learning objective combines the episode loss in Eq. 4 with the global loss in Eq. 7.
\begin{equation}
L(\theta, \omega) =
\mathbb{E}_{p(\tau)}\left[
L_e^\tau(\theta) + \lambda L^{\tau}_{g}(\theta,\omega)\right]
\label{eq:total_loss}
\end{equation}
\noindent Here, $\lambda$ is the hyperparameter for loss balancing and we simply use $\lambda=1$. In order to compute the final objective, we sample a single task and then average loss according to the task distribution $p(\tau)$ during training. This combined objective allows our model to match imbalance length pairs, while these pairs are classified over whole training classes together.
\begin{table}[t!]
\centering
\small
\caption{Verification performance on full utterance. C: Contrastive loss; A: A-softmax; SM: Softmax; NS: Normalized softmax; M: Meta-learning;  Spec: Spectrogram; *: With data augmentation; D: Development set.}
\vspace{-0.05in}
\resizebox{\linewidth}{!}{
 \begin{tabular}{c c c c c}
 \hline
 Model & 
 \multirow{2}{*}{M} &
 \multirow{2}{*}{Feature} & Train  &
 \multirow{2}{*}{$C_{det}^{min}$ / EER\%} \\
 (Aggregation+Loss) & & & dataset & \\
 \hline
 \hline
 i-vectors+PLDA \cite{nagrani2017voxceleb} &
 $\xmark$ &
 -- &
 Vox1(D) & 0.73 / 8.8\\
 VGG-M (TAP+C) \cite{nagrani2017voxceleb}  &
 $\xmark$ &
 Spec-512 & 
 Vox1(D) & 0.71 / 7.8\\
 ResNet34 (SAP+A) \cite{cai2018exploring} &
 $\xmark$ &
 MFB-64 & 
 Vox1(D) & 0.622 / 4.40\\
 ResNet34 (SPE+A) \cite{jung2019spatial} &
 $\xmark$ &
 MFB-64 &
 Vox1(D) & \textbf{0.402} / 4.03\\
 TDNN (ASP+SM) \cite{okabe2018attentive} &
  $\xmark$ &
 MFCC-40 &
 Vox1(D)* & 0.406 / 3.85\\
 \textbf{ResNet34 (TAP+NS)} &
  $\cmark$ &
 MFB-40 &
 Vox1(D) & 0.418 / \textbf{3.81}\\
 \hline
 UtterIdNet (TDV+SM) \cite{hajavi2019deep} &
  $\xmark$ &
 Spec-257 & Vox2(D) & -- / 4.26 \\
 Thin ResNet34 \cite{xie2019utterance}&
  \multirow{2}{*}{$\xmark$} &
 \multirow{2}{*}{Spec-257}&
 \multirow{2}{*}{Vox2(D)} & \multirow{2}{*}{-- / 3.22} \\
 (GhostVlad+SM) &  \\
 ResNet34 (SPE+A) \cite{jung2019spatial} &
 $\xmark$ &
 MFB-64 &
 Vox2(D) & 0.245 / 2.61 \\
 \textbf{ResNet34 (TAP+NS)} &
 $\cmark$ &
 MFB-40 &Vox2(D) & \textbf{0.234} / \textbf{2.08} \\
 
 \hline
 \end{tabular}
}
 \vspace{-0.20in}
 \label{tbl:full_comparison}
\end{table}

 
\begin{table*}[t!]
\normalsize
\centering
\small
\vspace{-0.27in}
    \caption{Verification performance on short utterance. G: Global classification; M: Meta-learning; D: Development set; T: Test set.}
    \vspace{-0.1in}
{\footnotesize
 \begin{tabular}{c c c c c c | c c c}
 \hline
 \multirow{2}{*}{Model (Aggregation+Loss)} & \multirow{2}{*}{G} & \multirow{2}{*}{M} & \multirow{2}{*}{Feature} &
  Train & Test &EER\% & EER\% & EER\% \\
  & & & & dataset & dataset & 1s & 2s & 5s\\
 \hline
 \hline
  ResNet34 (TAP+NS) & $\cmark$ & $\xmark$& MFB-40
  & Vox1(D) & Vox1(T) & 9.41 & 6.76 & 5.22 \\
  ResNet34 (TDV+NS) \cite{hajavi2019deep} & $\cmark$ & $\xmark$& MFB-40
  & Vox1(D) & Vox1(T) & 9.12 & 6.44 & 4.92 \\
  ResNet34 (TAP+NS) & $\xmark$& $\cmark$ & MFB-40
  & Vox1(D) & Vox1(T) & 8.44 & 6.33 & 4.81 \\
  \textbf{ResNet34 (TAP+NS)} & $\cmark$ & $\cmark$ & MFB-40
  & Vox1(D) & Vox1(T) & \textbf{7.53} & \textbf{5.39} & \textbf{4.03} \\
 \hline
 Thin ResNet34 (GhostVlad+SM) \cite{xie2019utterance} &  $\cmark$ & $\xmark$& Spec-257
  & Vox2(D) & Vox1(D+T) &
 12.71 & 6.59 & 3.34 \\
 ResNet34 (SAP+AM) \cite{gusev2020deep} & $\cmark$ & $\xmark$& MFB-80
  & Vox2(D)* & Vox1(D+T) & 9.91 & 4.48 & 2.26 \\
 \textbf{ResNet34 (TAP+NS)} & $\cmark$ & $\cmark$ & MFB-40
  & Vox2(D) & Vox1(D+T) &
  \textbf{5.31} & \textbf{3.15} & \textbf{2.17} \\
 \hline
 \end{tabular}
}
 \vspace{-0.15in}
    \label{tbl:short_duration}
\end{table*}


\begin{table}[t!]
\centering
\small
\vspace{-0.05in}
\caption{Performance comparison on length pairs of training utterances. All models are meta-learned without global classification. VoxCeleb1 development set and test set are used for training and testing, respectively.}
\vspace{-0.05in}
{\footnotesize
 \begin{tabular}{c c | c c c c}
 \hline
 Support & Query &
 EER\% & EER\% & EER\% & EER\% \\
 length & length & 1s & 2s & 5s & full\\
 \hline
 2s & 2s & 9.32 & 6.66 & 5.05 & 4.64\\
 2s & 1s & 8.81 & 6.56 & 5.36 & 4.94\\
 2s & 1s-2s & \textbf{8.44} & \textbf{6.33} & \textbf{4.81} & \textbf{4.52}\\

\hline
 \end{tabular}}
 \vspace{-0.25in}
 \label{tbl:ablation_length}
\end{table}
\vspace{-0.1in}
\section{Experiments}

\subsection{Dataset} We experiment our method under various settings on VoxCeleb datasets. VoxCeleb1 \cite{nagrani2017voxceleb} and VoxCeleb2 \cite{chung2018voxceleb2} are large scale text-independent speaker recognition datasets, each of which consists of 1251 and 5994 speakers, respectively. The two datasets have disjoint sets of speakers. We measure the speaker verification results with equal error rate (EER) and the minimum detection cost function (minDCF or $C^{min}_{det}$) at $P_{target}$ = 0.01. We score the veriﬁcation trials using cosine similarity. For unseen speaker identification, we report the average accuracy over 1000 randomly generated episodes with 95\% confidence intervals.

\subsection{Experiment setting}
We use 40-dimensional log mel-ﬁlterbank (MFB) features with a frame-length of 25 ms as input features, which overlaps the adjacent frames by 15ms. We mean-normalized the inputs along the time-axis without any voice activity detection (VAD) or data augmentation. In training episodes, we perform 1-shot 100-way classification, while setting the number of query examples for each class to 2. For memory efficiency, we set length of support set to 2 seconds and the length of the query to between half and full of the support length. For vanilla training, we use fixed length speech of 2 seconds. For frame-level feature extraction, we use ResNet34 with 32-64-128-256 channels for each residual stage. Extracted features are aggregated with TAP and are passed through the fully-connected layer to obtain 256-dimensional embeddings. We use SGD optimizer with the Nesterov momentum of 0.9 and set the weight decay to 0.0001. We set initial learning rate to 0.1 and decay it by a factor of 10 until convergence. Every experiment is done with a single NVIDIA 2080Ti GPU.

\subsection{Speaker verification for full utterance}
We first examine the results of full-duration SV to analyze the advantage of our training scheme. The results in Table \ref{tbl:full_comparison} show the model performances evaluated on VoxCeleb1 \cite{nagrani2017voxceleb} original test trial. For fair comparison, we report baselines without VAD and data augmentation except for x-vector \cite{snyder2018x} based models \cite{okabe2018attentive}. On VoxCeleb1, our proposed model outperforms previous state-of-the-arts models. For the same backbone (i.e. ResNet34), our model achieves superior performance without any aggregation and margin-based metrics. In general, additional aggregation and margin-based metric lead to the better performance. Further, our model outperforms time delay neural network (TDNN) with attentive statistic pooling \cite{okabe2018attentive}. Our model also consistently outperforms baseline models, on a larger VoxCeleb2 \cite{chung2018voxceleb2} dataset. 

\subsection{Speaker verification for short utterance}
We first describe the experimental settings, then report the results of our model and other previous state-of-the-arts models for short utterance. We test our model on two datasets: 1) The original VoxCeleb1 test trial which is the same datset used to evaluate full utterances, and 2) VoxCeleb1 full dataset (1251 speakers in total). We use full-duration enrollment utterances, but randomly cropped the test utterances by 1, 2 and 5 seconds. If a test utterance is shorter than required we set it to the target length by duplicating its own segment. 

To show the efficacy of our method, we perform an ablation study with VoxCeleb1 on upper rows of Table \ref{tbl:short_duration}. We observe that TDV \cite{hajavi2019deep} which aims at short segments outperforms temporal average pooling with slight margin. However, the result in the third row shows that model only trained with meta-learning outperforms TDV and conventionally trained model (See first row). Further, our proposed model which combines meta-learning with global classification obtains the best performance against other baselines trained on VoxCeleb1 with large margin.

For the comparison against other previous state-of-the-arts models \cite{hajavi2019deep, xie2019utterance, gusev2020deep}, we trained the model with VoxCeleb2 dataset and tested on VoxCeleb1 full dataset. We use the same trial as described in \cite{gusev2020deep}. For every speaker, we randomly generate trials for 100 positive pairs and 100 negative pairs. Bottom rows in Table \ref{tbl:short_duration} show that our model outperforms baseline models with significantly large margin for 1-2 seconds. Since our model does not use any aggregation techniques or margin-based optimization, we can say that its impressive improvement mostly comes from our combined learning scheme. Furthermore, note that our model uses only 40-dimensional features but the baselines use features with more than twice the dimensions. Thus our model may obtain larger performance if we use higher dimensional inputs. For the comparison against \cite{hajavi2019deep}, since UtterIdNet is not publicly available, we compared it using TDV instead.

Our performance gain is due to two reasons. First, we compose training episode with imbalance length pairs, where utterance length of query largely varies and is shorter than the length of the utterances in the support set. In Table \ref{tbl:ablation_length}, we observe that the proposed imbalance length pairs setting outperforms both equal length pairs and fixed long-short pairs. Note that \cite{wang2019centroid, anand2019few} use equal-length pairs. In our proposed setting, model comes across various length pairs at each episode, and then is meta-learned such that it can well-match the imbalanced length pairs and become robust to speech duration. Secondly, to learn more discriminative embeddings, we classify both the support and the query samples against the entire set of training classes. Unlike the conventional method which classifies the utterance of same length for each batch, our combined scheme classifies utterances of different lengths at once. It results in reduction of variance caused by speech duration and enhances inter-class clustering. By combining these two components, our proposed model achieves the state-of-the-arts performance on short utterances, while yielding good performance on full utterances.

\begin{table}[t!]
\centering
\small
\vspace{-0.05in}
\caption{Accuracy (\%) of unseen speaker identification.}
\vspace{-0.1in}
\resizebox{\linewidth}{!}{
 \begin{tabular}{c | c | c c c c}
 \hline
 Query & Training & 
 \multirow{2}{*}{5-way} & \multirow{2}{*}{20-way} & \multirow{2}{*}{50-way} &
 \multirow{2}{*}{100-way} \\
 length & method & 
  &  &  &  \\
 \hline
 \multirow{2}{*}{1s} & Vanilla
 & 94.77\tiny$\pm$0.33 & 85.63\tiny$\pm$0.29 & 77.72\tiny$\pm$0.22 & 70.76\tiny$\pm$0.17\\
 & \textbf{Ours}
  & \textbf{96.40\tiny$\pm$0.30} & \textbf{88.92\tiny$\pm$0.25} & \textbf{82.10\tiny$\pm$0.20} & \textbf{75.92\tiny$\pm$0.16}\\
 \hline
 \multirow{2}{*}{2s} & Vanilla
 & 97.18\tiny$\pm$0.27 & 92.18\tiny$\pm$0.23 & 86.63\tiny$\pm$0.19 & 81.46\tiny$\pm$0.15\\
 & \textbf{Ours}
 & \textbf{98.38\tiny$\pm$0.20} & \textbf{94.91\tiny$\pm$0.19} & \textbf{90.90\tiny$\pm$0.16} & \textbf{86.90\tiny$\pm$0.13}\\
\hline
 \multirow{2}{*}{5s} & Vanilla
 & 98.27\tiny$\pm$0.24 & 94.95\tiny$\pm$0.18 & 91.30\tiny$\pm$0.16 & 87.61\tiny$\pm$0.13\\
 & \textbf{Ours}
 & \textbf{99.12\tiny$\pm$0.16} & \textbf{97.15\tiny$\pm$0.14} & \textbf{94.88\tiny$\pm$0.12} & \textbf{92.34\tiny$\pm$0.10}\\
\hline
 \end{tabular}
}
 \vspace{-0.15in}
 \label{tbl:unseen_SI}
\end{table}

\subsection{Unseen speaker identification}
We now evaluate the performance of our model on unseen speaker identification tasks. To analyze our model, we trained the model on VoxCeleb2 dataset and tested it on the whole VoxCeleb1 dataset. As done in the verification experiments, we enroll with one utterance for each speaker and set the enrollment utterances equally to 5 seconds. Therefore, we randomly sample $N$-speakers from VoxCeleb1 dataset, and then sample 1 and 5 utterances from each speaker for enrollment and test utterance, respectively. For utterance shorter than required, we handled it as done in 4.4. As shown in Table \ref{tbl:unseen_SI}, our proposed method outperforms vanilla training in every setting. The performance gap increases as the number of speakers grows. Generally, the performance of identification decreases as the number of speakers becomes larger and the length of utterance becomes shorter.
\section{Conclusion}
We proposed a novel meta-learning scheme for short duration speaker recognition. In order to simulate practical settings in training, we propose an episode composition in which the support and query set have different speech lengths, and combined the meta-learning scheme with global classification for obtaining discriminative embedding space. We validate our model on various speaker recognition tasks on VoxCeleb datasets, and obtain the state-of-the-art performances on short utterance speaker recognition.
\section{Acknowledgements}

This material is based upon work supported by the Ministry of Trade, Industry and Energy (MOTIE, Korea) under Industrial Technology Innovation Program (No.10063424, Development of distant speech recognition and multi-task dialog processing technologies for in-door conversational robots).

\bibliography{ref}
\bibliographystyle{IEEEtran}

\end{document}